\documentclass[prl,aps,twocolumn,showpacs,notitlepage,footinbib]{revtex4-1}

\usepackage{hyperref}
\usepackage{enumitem}
\usepackage{graphicx}
\usepackage{amsmath}
\usepackage{amsfonts}
\usepackage{amssymb}
\usepackage{times}
\usepackage{color}
\usepackage{bbm}
\usepackage{bm}
\usepackage{xcolor}
\usepackage[export]{adjustbox}
\usepackage[colorinlistoftodos, shadow,textsize= footnotesize]{todonotes}
\usepackage[caption=false]{subfig}

\newcommand{\ket}[1]{\displaystyle{|#1\rangle}}

\newcommand{\be}{\begin{equation}}
\newcommand{\ee}{\end{equation}}

\newcommand{\beq}{\begin{eqnarray}}
\newcommand{\eeq}{\end{eqnarray}}

\begin{document}

\title{Nonlocality with ultracold atoms in a lattice}
\author{Sophie Pelisson} 
\affiliation{QSTAR, INO-CNR and LENS, Largo Enrico Fermi 2, I-50125 Firenze, Italy}
\author{Luca Pezz\`e}
\affiliation{QSTAR, INO-CNR and LENS, Largo Enrico Fermi 2, I-50125 Firenze, Italy}
\author{Augusto Smerzi}
\affiliation{QSTAR, INO-CNR and LENS, Largo Enrico Fermi 2, I-50125 Firenze, Italy}
\date{\today}

\begin{abstract}
We study the creation of nonlocal states with ultracold atoms trapped in an optical lattice. We show that these states violate Bell inequality by measuring one- and two-body correlations. 
Our scheme only requires beam splitting operations and global phase shifts, and can be realized within the current technology, 
employing single-site addressing. 
This proposal paves the way to study multipartite nonlocality and entanglement in ultracold atomic systems.
\end{abstract}


\maketitle

Bell's theorem~\cite{Bell87} unveils the incompatibility between quantum mechanics 
and a generic local hidden variable theory via a set of inequalities that can be experimentally tested. 
The notion of Bell nonlocality  (see~\cite{reviews} for a recent review) 
has been originally proposed and mainly developed in the case of two parties. 
It is now recognized as a crucial resource for several quantum information processing, 
as secure communication~\cite{BuhrmanRevModPhys10,EkertPRL91} and certified random number generation~\cite{PironioNature10}. 
Violations of Bell's inequalities have been reported in a variety of bipartite physical systems, giving a strong evidence that nature is nonlocal \cite{notaBell}. 
Experiments have been mostly performed with photons~\cite{AspectPRL82, OuPRL88,RarityPRL90,WeihsPRL1998,GroblacherNATURE2007} 
and, in recent years, also with ions~\cite{RoweNATURE2001, SakaiPRL06, MatsukevichPRL2008} and atom-photon hybrid systems~\cite{MoehringPRL2004, MatsukevichPRL2005}.
Quantum nonlocality in many-partite systems is, by far, more complex and less developed than in the bipartite case.
One crucial obstacle is that multipartite Bell's inequalities~\cite{WernerPRA01,ZukowskiPRL02} usually involve high-order 
correlations that should be measured for $N$ parties in order to demonstrate nonlocality. 
This poses serious experimental -- but also theoretical -- challenges for large $N$.
Presently, the experimental violation of multipartite Bell's inequalities has been achieved
with three~\cite{PanNATURE2000, EiblPRL2004, LavoieNJP2009} and four~\cite{EiblPRL2003, ZhaoPRL2003} 
photons and trapped ions~\cite{LanyonPRL2014}, whereas most of these are limited to GHZ states.
Much less is known about the possibility to investigate quantum nonlocality for ultracold neutral atoms. 
These systems offer the practical advantage of large detection efficiencies and a variety of control techniques, including single site addressing and  
fine manipulation of internal and external degrees of freedom.
Recent progresses have evidenced that ultracold atoms are optimal candidates 
for the creation of entangled states with a large number of particles~\cite{EsteveNature08,RiedelNature10,GrossNature10,LuckeScience11,HaasSCIENCE2014}.
However, while entanglement is generally a prerequisite for the violation of Bell's inequalities~\cite{note01}, 
to date, there are only few proposals~\cite{MullinPRA08, GneitingPRL2008, Lewis-SwanARXIV} to demonstrate quantum nonlocality in these many-body systems.


\begin{figure}[t!]
\includegraphics[width=1\columnwidth]{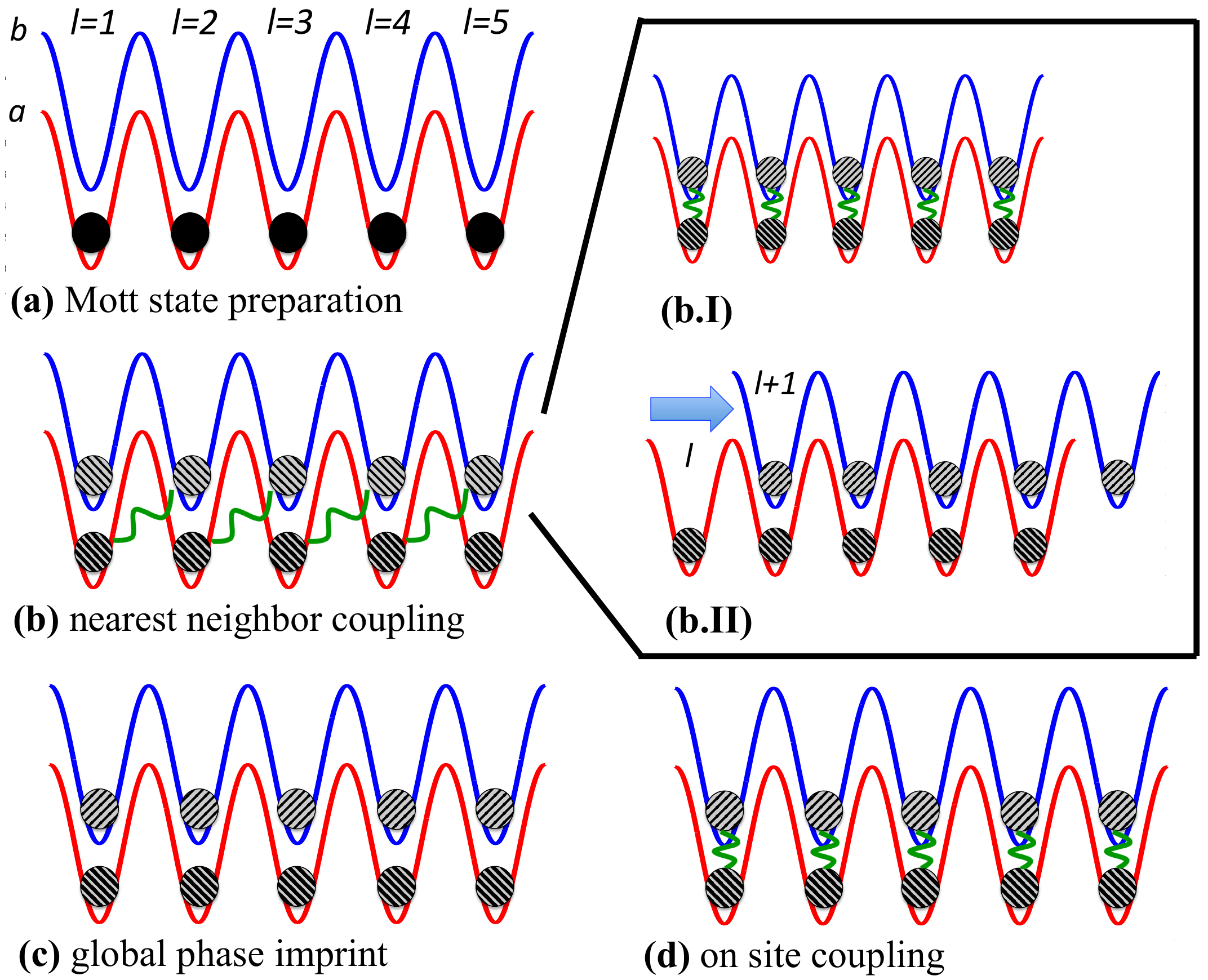}
\caption{(color online) 
Quantum nonlocality of neutral atoms in an optical lattice (here limited, for illustration sake, to few lattice sites). 
The protocol consists of four operations involving internal (hyperfine) states and it is implemented by a state-selective optical lattice.
(a) Preparation of a system of two-level atoms ($a$ and $b$ indicate hyperfine levels, $l$ the lattice site) with one atom per lattice site, 
each atom being schematically represented by a dot.
(b) Balance pulse coupling neighboring wells: this is made of a $\pi/2$ Rabi pulse, schematically represented by a green waving line, 
(b.I) followed by a relative shift of the state-selective lattice (b.II).  
(c) Collective phase shift, $\theta$ of $\varphi$, see text; 
(d) Balance pulse coupling the internal levels of each atom.
The protocol ends with the measurement of the number of particles in the internal ground level.}
\label{lattice} \label{fig1}
\end{figure}


Here we propose a realistic experimental protocol to create and observe Bell's nonlocality of an arbitrary 
large number of neutral atoms trapped in an homogeneous one-dimensional optical lattice. 
The crucial ingredient of the proposal is the coupling between hyperfine states of atoms sit at neighboring lattice sites.
This coupling can be accomplished by resonant light interaction and a moving state-dependent lattice potential, as outlined in Fig.~\ref{lattice}.
We notice that the combination of internal Rabi splitting and the coherent transport of atoms in a spin-dependent optical lattice has been 
first experimentally demonstrated in \cite{MandelPRL2003}.
This capability has been further used to generate entangling gates for neutral atoms \cite{MandelNATURE2003, AnderliniNATURE2007},  
super-exchange interaction \cite{TrotskiSCIENCE2008}, quantum random walks \cite{KarskiSCIENCE2009}, and spin-squeezed states \cite{RiedelNature10}.   
Our protocol for the violation of Bell inequalities consists of four steps. 
A Bose gas is initially prepared with one atom per lattice site, see Fig.~\ref{fig1}(a), each atom having two internal levels indicated as $a$ and $b$.
This can be accomplished by starting from a superfluid gas and then raising the potential barriers in such a way to 
enter the Mott-insulator phase \cite{GreinerNATURE2002}, eventually suppressing tunneling between neighboring wells.
The system is thus prepared in the state $\ket{\psi_0} =\bigotimes_l \ket{a}_l$, where $\ket{k}_l$ indicates an atoms in the internal state $k=a,b$ at lattice site $l=1,2,..., N$.
As second step, 
each atom is placed in the coherent superposition $\ket{a}_l  \to (\ket{a}_l + \ket{b}_{l+1})/\sqrt{2}$, see Fig.~\ref{fig1}(b).
This operation (linear in the $N$ particles) creates correlations between lattice sites, which are our parties. 
It will be shown that these correlations are nonlocal as they are responsible for the violation of Bell's inequalities. 
Nearest neighbor coupling can be experimentally realized following two stages:
{I)} a $\pi/2$ Rabi pulse for atoms at each lattice site, $\ket{a}_l  \to (\ket{a}_l + \ket{b}_{l})/\sqrt{2}$ [Fig.~\ref{fig1}(b,I)];  
followed by {II)} a shift of the state-dependent lattice potential $\ket{b}_l \to \ket{b}_{l+1}$ [Fig.~\ref{fig1}(b,II)] that can be realized by 
changing the polarization angle of two linearly-polarized and counter-propagating laser fields \cite{JakschPRL1999, MandelPRL2003}. 
As third step, we apply a local phase shift $e^{i \theta_l \hat{n}_{g,l}}$ at each lattice site [Fig.~\ref{fig1}(c)], 
where $ \hat{n}_{g,l}$ is the number operator and 
 $\theta_l$ are a set of local phases. 
Below we show that the optimal choice for the phase shifts involves imprinting the same phase to each atom. 
This is an important simplification for the experimental implementation of the protocol. 
Phase imprinting should happen on a time scale much faster than the interaction time scale between the $\ket{g}_l$ and $\ket{e}_{l+1}$ atoms.
We point out that, differently from the creation of entangling gates \cite{JakschPRL1999, SorensenPRL1999}, 
our protocol does not involve interaction between atoms.
The final operation is an on-site coupling pulse $\ket{a}_l \to (\ket{a}_l + \ket{b}_{l})/\sqrt{2}$ and $\ket{b}_l \to (\ket{b}_l - \ket{a}_{l})/\sqrt{2}$  [Fig.~\ref{fig1}(d)], 
formally described as $e^{-i\frac{\pi}{2}J_y}$ with  
$J_y = \tfrac{1}{2} \sum_{l=1}^N  \hat{\sigma}_y^l$ and $ \hat{\sigma}_y^{(l)}$ the Pauli matrix for the $l$th site.
The protocol ends by counting the total number of particles in the internal ground state $N_g={\sum\limits_{l=1}^N n_{g,l} }$, 
from which we obtain a dichotomic measurement given by the parity $P_g=(-1)^{N_g}$. 
We emphasize that the whole protocol 
does not involve vibrational states of the well.

Our criterion for witnessing quantum nonlocality of the ultracold atomic system is based on the violation of a novel class 
of multipartite Bell's inequalities, introduced in \cite{TuraScience14} and involving only one- and two-body correlations. 
These inequalities are 
\begin{equation}
B_N(\theta_i,\varphi_i) =\alpha \Big(S_0+ \frac{S_1}{N} \Big)+\frac{\gamma}{2}S_{00}+\frac{N}{2} S_{01}-\frac{S_{11}}{2}\geq -\beta_c^{(N)},
\label{Tura}
\end{equation}
where 
\begin{equation}
S_k=\sum\limits_{i=1}^N\langle\mathcal{M}_k^{(i)}\rangle \qquad S_{kl}=\sum\limits_{\substack{i,j=1\\i\neq j}}^N\langle\mathcal{M}_k^{(i)}\mathcal{M}_l^{(j)}\rangle
\end{equation}
denote the one- and two-body correlations, with $\mathcal{M}_k^{(i)}$ ($k=0,1$) representing the two different local measurement observables realized in well $l$:
\begin{equation}\begin{split}
\mathcal{M}_0^{(l)}&=e^{-i\theta_l n_{g,l}}e^{i\frac{\pi}{2}J_y}(-1)^{N_g}e^{-i\frac{\pi}{2}J_y}e^{i\theta_l n_{g,l}}, \\
\mathcal{M}_1^{(l)}&=e^{-i\varphi_l n_{g,l}}e^{i\frac{\pi}{2}J_y}(-1)^{N_g}e^{-i\frac{\pi}{2}J_y}e^{i\varphi_l n_{g,l}}.
\label{meas}\end{split}\end{equation} 
The two measurements differ by local phase shifts, $\theta_{l}$ and $\varphi_{l}$, applied to the ground state. 
These correspond to the two different detector settings in the Clauser-Horne-Shimony-Holt (CHSH) scheme \cite{ClauserPRL69}. 
We choose $\alpha= N(N-1)\left(\lceil\frac{N}{2}\rceil-\frac{N}{2}\right)$ and $\gamma=\frac{N(N-1)}{2}$,
in accordance with the parameters used in \cite{TuraScience14} to trace nonlocality of Dicke states, 
for which the classical limit is set by $\beta_c^{(N)}=\frac{N(N-1)}{2}\lceil\frac{N+2}{2}\rceil$.
To illustrate our proposal, we will first focus on the simple case of two atoms trapped in two wells and then generalize the discussion to many 
atoms and many wells.


{\it Two-wells case.}
For two and three wells our proposal reduces to the scheme first proposed by Yurke and Stoler~\cite{YurkePRA92,YurkePRL92} for optical systems. 
In the double-well case, the final state is $\vert \psi_{\rm fin} \rangle = ( e^{i (\theta_1 + \theta_2)/2} \vert \psi_1 \rangle + \sqrt{2} \vert \psi_2 \rangle )/4$, 
where 
\begin{eqnarray}
\ket{\psi_1} &= & c \left(\ket{1010}+\ket{0101}\right) +i s \left(\ket{1001}+\ket{0110}\right),  \\
\vert \psi_2 \rangle & =& e^{i\theta_1}\left(\ket{0200}-\ket{2000}\right)+ e^{i\theta_2}\left(\ket{0002}-\ket{0020}\right), \quad
\label{state2W}\end{eqnarray}  
$c = \cos \tfrac{\theta_1+\theta_2}{2}$, $s = \sin \tfrac{\theta_1+\theta_2}{2}$ and 
$\ket{n_{g,1} n_{e,1} n_{g,2} n_{e,2}}$ represents the state of occupation numbers in each mode of the system, $0 \leq n_{g,l} \leq 2$.
From Eq.~(\ref{state2W}), we can obtain analytically the probability of the possible measurement results.
Similarly to Ref.~\cite{YurkePRA92}, let us denote with $P(\epsilon_1,\epsilon_2)$ the probability that an 
event $\epsilon_m$ occurs in the well $m$. 
As we consider only bosonic atoms, these events are elements of the set $[0,G,E,G^2,E^2]$, where $G$ represents 
one particle in its ground state and $E$ one particle in its excited state. 
We can distinguish three subsets of events obtained by the measurement of the population of each atomic state. 
These subset are: firstly the case where both the two wells are populated with atoms in the same internal state $A=\{GG,EE\}$, the second one represents the case where both the two wells are populated but with atoms in different internal states $B=\{GE,EG\}$ and finally, we have the case where only one well is populated with the two atoms in the same state $C=\{G^20,E^20,0G^2,0E^2\}$. 
The corresponding probabilities are:
\begin{equation}
P(\epsilon_1,\epsilon_2)=
\left\{\begin{array}{ll}
\tfrac{1}{4}\cos^2(\frac{\theta_1+\theta_2}{2}) \qquad &\text{if}\quad \epsilon_1,\epsilon_2\in A,\\[1mm]
\tfrac{1}{4}\sin^2(\frac{\theta_1+\theta_2}{2}) \qquad &\text{if}\quad \epsilon_1,\epsilon_2\in B,\\[1mm]
\tfrac{1}{8}\qquad &\text{if}\quad \epsilon_1,\epsilon_2\in C.
\end{array}\right.\end{equation}
We can see here that independently of the local phases, if one measures the atomic population in one well, one can predict whether the total results belongs to the set $A\bigcup B$ or $C$. 
This is coherent with local realism as it is a simple consequence of the conservation of the total number of particles in the system. 
As a consequence, we can directly ignore the part of the state belonging to the set $C$ and test the nonlocality for the reduced state $\vert \psi_1 \rangle / 2$ 
which we propose to study using the criterion \eqref{Tura}. 
We stress here that for two particles the class of Bell's inequalities (\ref{Tura}) reduces to the well-known CHSH inequality ~\cite{ClauserPRL69}:
\begin{equation}
-2\leq\frac{S_{00}-S_{11}}{2}+S_{01}\leq2.
\label{CHSH}
\end{equation}
Using the set of measurement described previously leads to the following expression for the mean value of the Bell's operator $B_2(\theta_1,\theta_2,\varphi_1,\varphi_2)$:
\begin{equation}\begin{split}
B_2(\theta_1,\theta_2,\varphi_1,\varphi_2)=&\cos\left(\theta_1+\theta_2\right)-\cos\left(\varphi_1+\varphi_2\right)\\
&+\cos\left(\varphi_1+\theta_2\right)+\cos\left(\theta_1+\varphi_2\right).
\end{split}\end{equation}
The optimal choice for the phase shifts can be rewritten as $\theta_1+\theta_2=-(\varphi_1+\theta_2)=-(\theta_1+\varphi_2)=\omega$ and $-(\varphi_1+\varphi_2)=3\omega$ as in Ref.~\cite{MullinPRA08}. 
The expression \eqref{CHSH} becomes $-2\leq 3\cos(\omega)-\cos(3\omega)\leq2$ which is maximal for $\omega=\pi/4$ and we find $B_2^{(max)}\simeq2\sqrt{2}$. 
This violation corresponds to the maximal violation predicted by the Cirel'son's bound \cite{CirelsonLettMathPhys80}. 
It is worth to note here that a violation of the CHSH inequality is also observed without the reduction of the state \eqref{state2W} for the same set of values of $\theta_1$, $\theta_2$, $\varphi_1$ and $\varphi_2$ for which we then find a smaller violation i.e. $B_2^{(max)}\simeq2.41$ in accordance with Refs.~\cite{MullinPRA08} 
and~\cite{YurkePRA92}.

\begin{figure}[t!]\center
\includegraphics[width=.95\columnwidth]{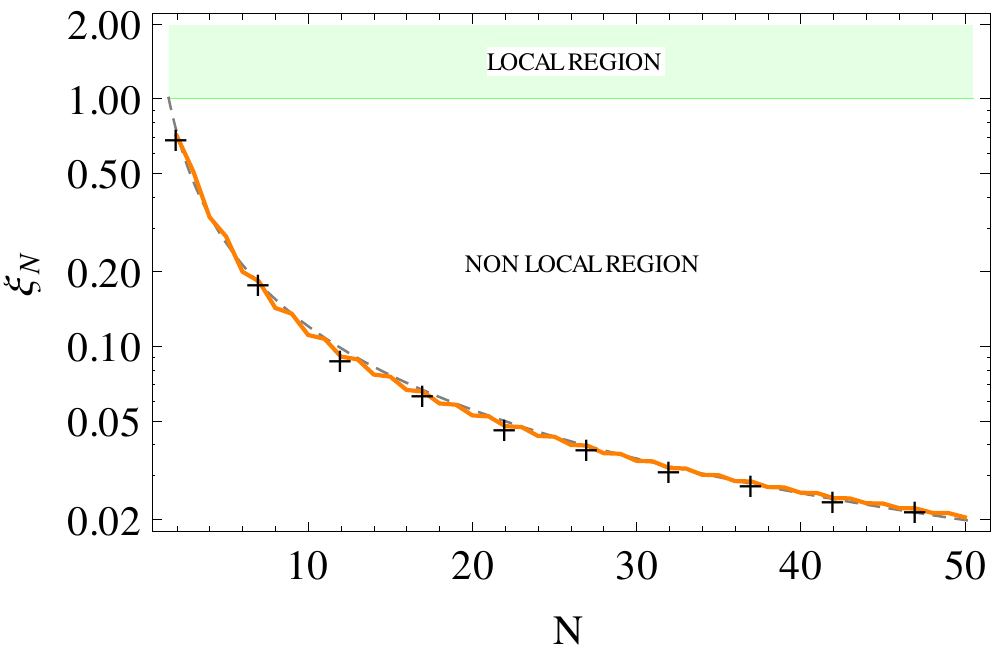}
\caption{\label{PlotBell}
(colors online) 
Maximal relative violation $\xi_N$ of Eq.~\eqref{Tura}, optimized over different local phases, 
as a function of the number of parties $N$ (crosses). 
The orange line is $\xi_N^{glob}$, where all local phases are equal, $\theta_k=\theta$ and $\varphi_k=\varphi$.
The dashed grey line is a power law fit $1.5/N$.}
\end{figure}
\begin{figure*}[!t]
\includegraphics[width=.48\columnwidth,valign=c]{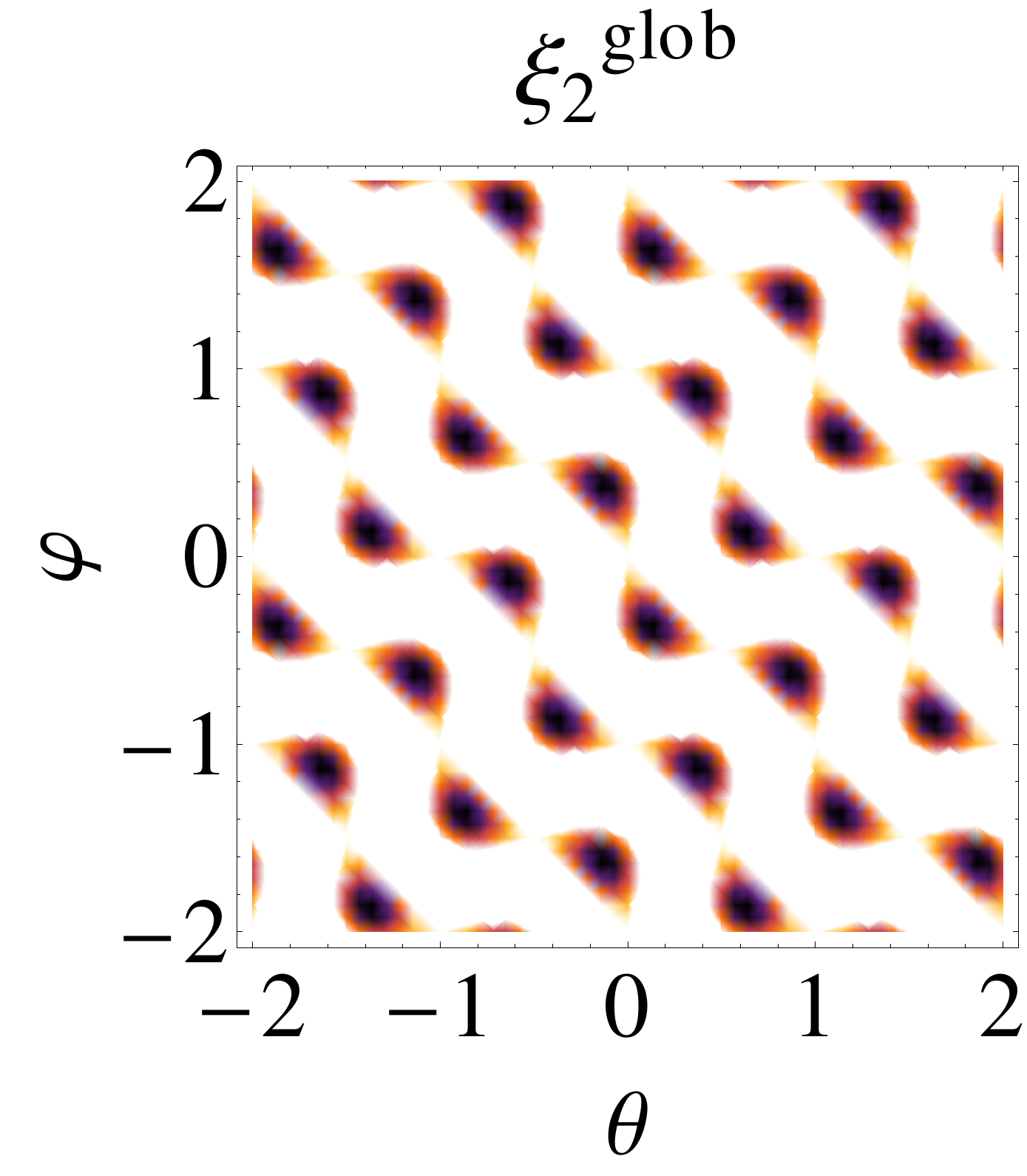}
\includegraphics[width=.48\columnwidth,valign=c]{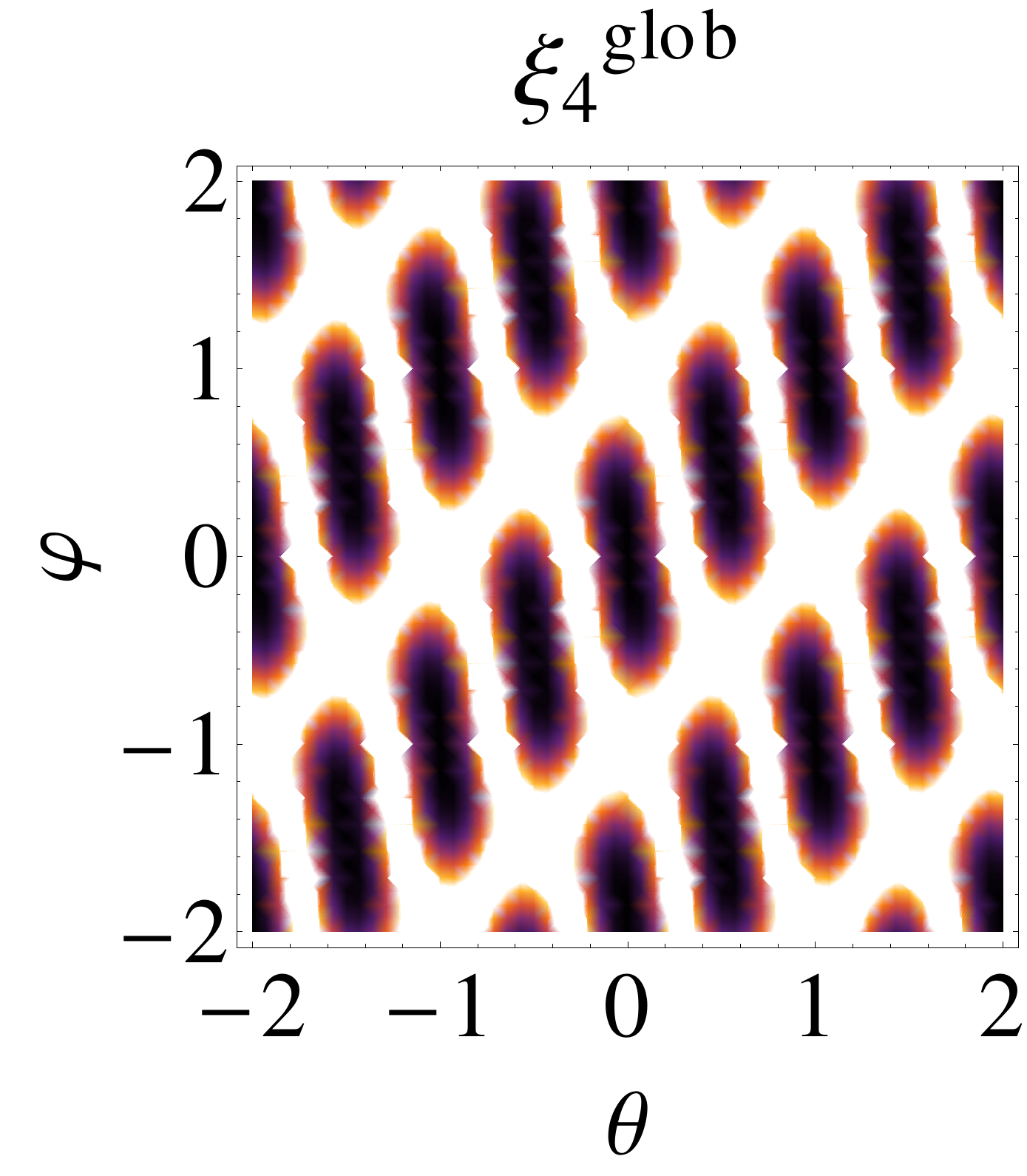}
\includegraphics[width=.48\columnwidth,valign=c]{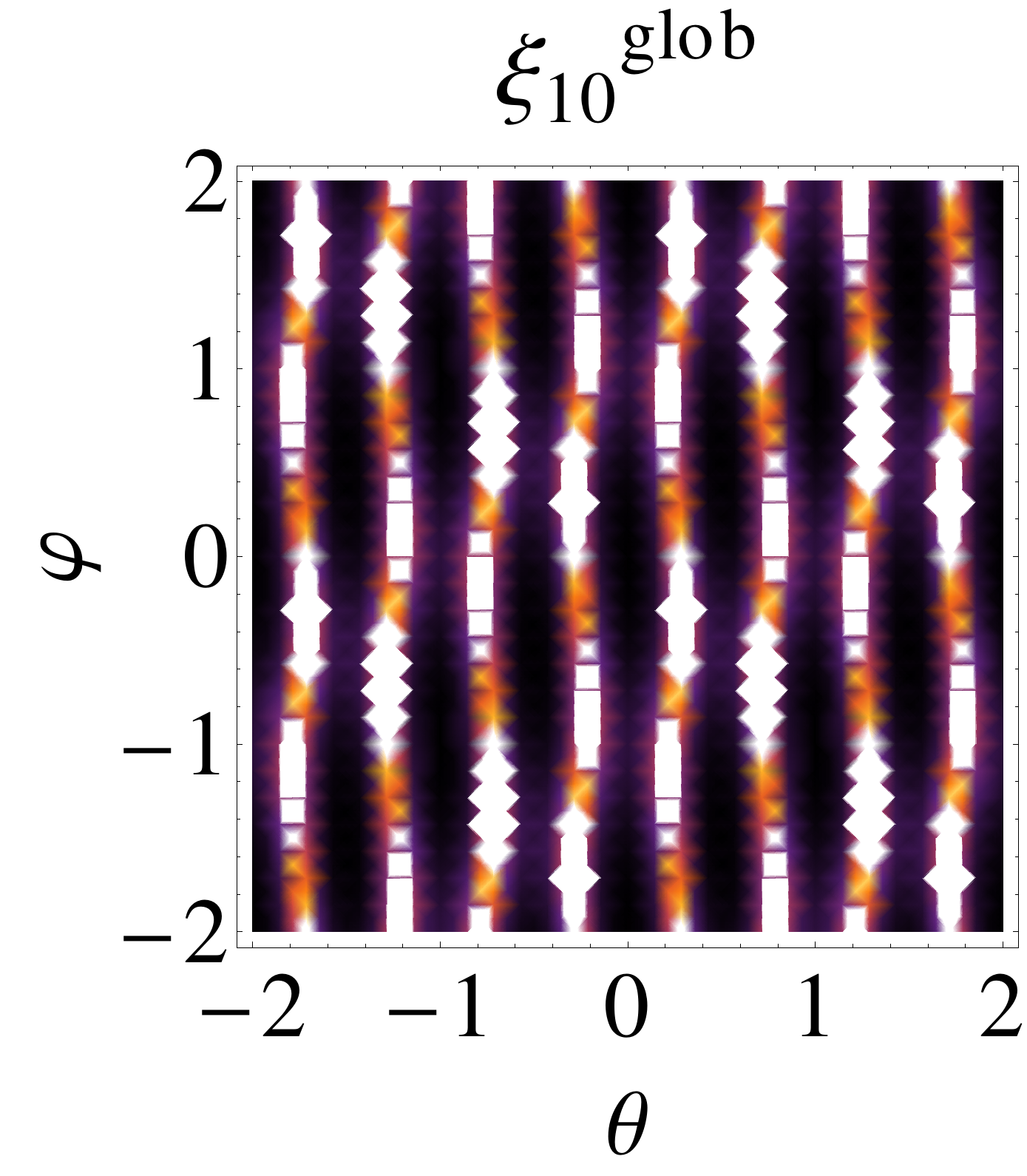}
\includegraphics[width=.48\columnwidth,valign=c]{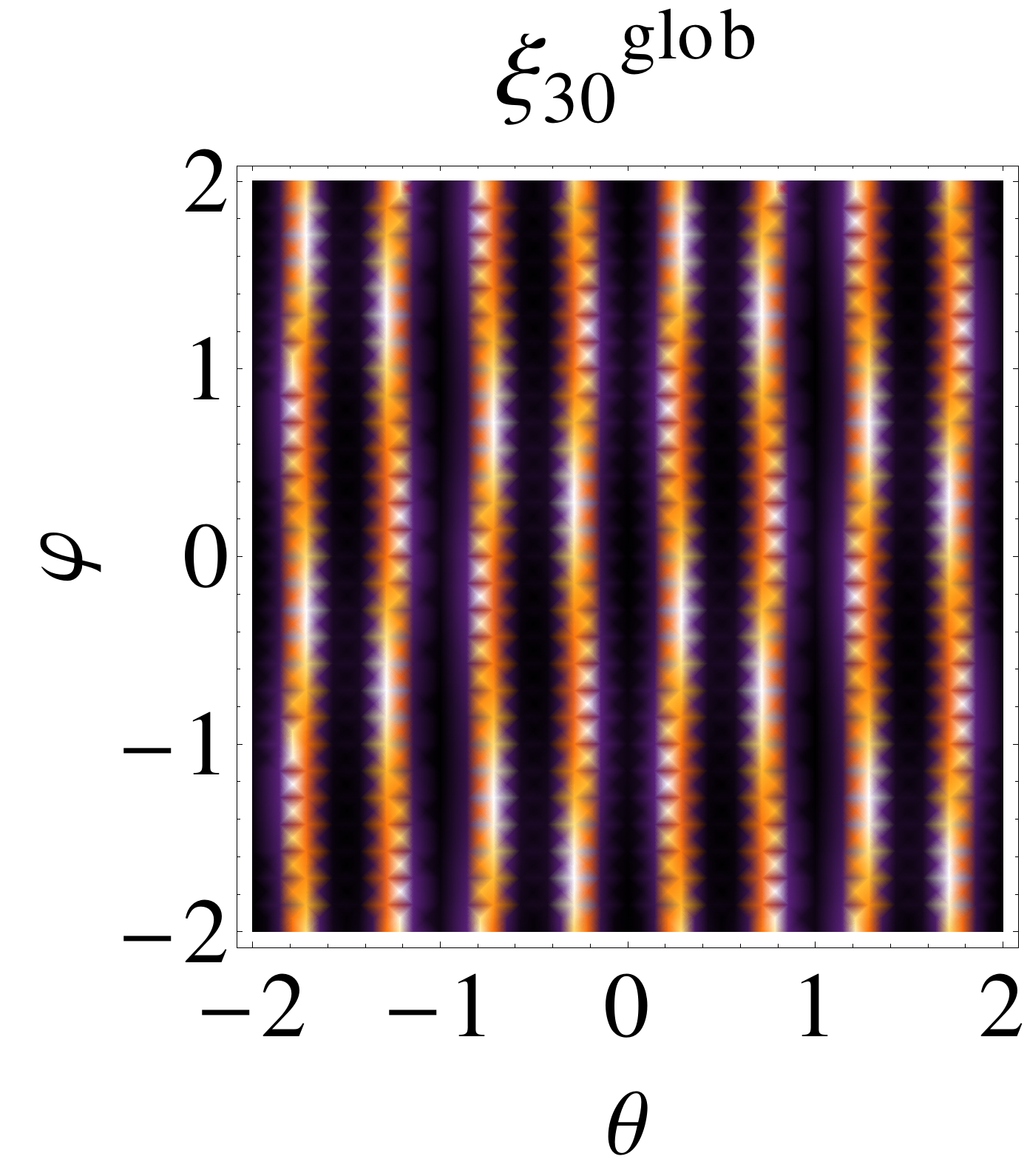}
\hspace{.2cm}\includegraphics[height=.408\columnwidth,valign=c]{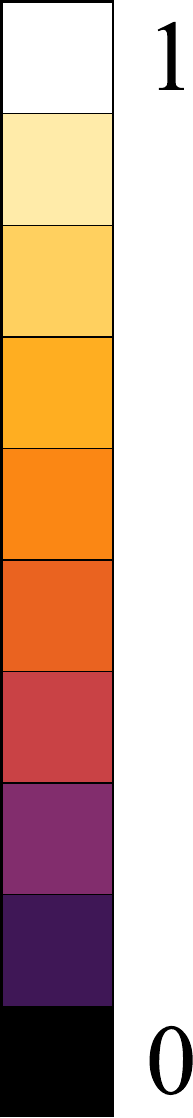}
\caption{(color online) Relative quantum violation $\xi_N^{glob}$ taking the same local phase shift, $\theta$ and $\varphi$, 
at each lattice site. Different panels refer to $N=2,4,10,30$ (from left to right).
The color scale is cut at $\xi_N^{glob} = 1$ and 
violation of Bell's inequality ($\xi_N^{glob} < 1$) is obtained in the colored regions. 
}
\label{density}
\end{figure*}

{\it Many-wells case.}
We now generalize the discussion to the many-well case. 
In particular, we emphasize that the case $N>2$ requires postselection of the output results and therefore single-site imaging of the optical lattice, 
a technique which has been successfully demonstrated experimentally \cite{BakrScience10}. 
In the two-wells case, post-selection of the state allows a stronger violation of the Bell's 
inequalities but is not compulsory to observe it and has not been considered in the previous section.
After the last beam-splitting operation of the protocol outlined in Fig.~\ref{lattice}, 
we obtain (see for instance Eq.~\eqref{state2W} for the double-well case) all wells populated by only one atom, with a probability $1/2^{N-1}$, or
at least one empty well, with a probability $1-1/2^{N-1}$. 
In the latter case, observing the number of particle in $N-1$ wells allows to know with certainty the population of the last one, 
independently of the local detectors settings $\{\theta_k,\varphi_k\}$. 
As a consequence, the correlations involving states with at least one well empty are classical correlations. 
A violation of Eq.~(\ref{Tura}) requires to ignore this case. 
It should be noticed that the same post selection has been considered and proved necessary elsewhere \cite{YurkePRL92,MullinPRA08}
in order to observe the GHZ contradiction with three particles \cite{GreenbergerAmJPhys90}.  
After post-selection, we can express the mean value of the Bell's operator in Eq.~\eqref{Tura} as
\begin{equation}\begin{split}
&B_N(\theta_k,\varphi_k) = \sum\limits_{k=1}^{N} \alpha\cos\theta_k+\beta\cos\varphi_k  + \\
&+\sum\limits_{\substack{k,l=1\\k\neq l}}^N \frac{\gamma}{2}\cos(\theta_k+\theta_l)+\delta\cos(\theta_k+\varphi_l)+\frac{\epsilon}{2}\cos(\varphi_k+\varphi_l).
\end{split}\end{equation}
We further minimize this expression as a function of $\theta_k,\varphi_k$ using a genetic algorithm~\cite{Haupt04}.
In Fig.~\ref{PlotBell}, we plot the maximum quantum violation of Eq. \eqref{Tura} normalized by the classical bound, 
$\xi_N \equiv  \beta_c^{(N)}/\min_{\theta_k, \varphi_k}\left[B_N(\theta_k, \varphi_k)\right]$, as a function of the number of parties $N$ of the system. 
The figure compares $\xi_N$ (crosses) with 
$\xi_N^{\rm glob} \equiv   \beta_c^{(N)}/\min_{\theta, \varphi}\left[B_N(\theta, \varphi)\right]$ (orange line)
obtained with a global optimization assuming that all the local phases are equal.
We thus conclude that the case $\theta_k=\theta$ and $\varphi_k=\varphi$ is very close to being an optimal choice of local phase.
This is of particular interest from the experimental point of view as it is far easier to shift the whole system with the same 
phase for each well instead of applying a different phase shift to each party. 
Furthermore, a simple fit of our results shows that $B_N \sim 1/N$, comparable to the result obtained in Ref. \cite{TuraScience14} for Dicke states. 
Figure \ref{density} shows the behavior of our relative quantum violation as a function of the phase shifts $\theta$ and $\varphi$ for different number of parties involved.
There, we plot only the values of 
$\xi_N^{\rm glob} < 1$,
{\it i.e.} the colored regions represent the region where we observe a violation. 
We see here that as the number of parties increases, the robustness of the violation against the fluctuations of the phase shifts $\theta$ and $\varphi$ increases as well. 


{\it Discussion.}
Having in sight the possibility to realize experimentally the proposed scheme, some technical issues must be studied further. 
First, let us emphasize that the above discussion, focused on Bose gases, can be easily generalized to fermions. 
Furthermore, we do not need to assume the particles originate from a common source (an initial superfluid, for instance):
interestingly the protocol works even if the particle have never seen each others \cite{YurkePRA92, YurkePRL92,MullinPRA08}.


A possible obstacle to the observation of nonlocality is given by interactions between atoms which can 
affect the first beam splitting operations.
To evaluate how the interactions impact the violation of the inequality \eqref{Tura}, we 
generalize the scheme of Fig.~\ref{fig1} replacing the linear beam-splitter operation $e^{-i \tfrac{\pi}{2} J_y}$ in Eq.~\eqref{meas}
by a non-linear one $U=e^{-i\frac{\pi}{2}\left(J_y+\chi J_z^2\right)}$ \cite{PezzePRA06}, 
where $J_z = \tfrac{1}{2}\sum_{l=1}^N  \hat{\sigma}_z^{(l)}$. 
The coefficient $\chi$ represents the relative strength of interaction with respect to the coupling between wells. 
We then estimate the role played by the interactions 
on the Bell operator $B_2(\theta_1,\theta_2,\varphi_1,\varphi_2)$ 
(for simplicity we consider only the two-wells case here), for optimized phase shifts. 
This calculation can be performed analytically for two wells~\cite{supp}. 
In the Rabi regime, $\chi \ll 1$, we obtain 
\begin{equation}
B_2\simeq2\sqrt{2}-\frac{\chi^2}{\sqrt{2}},
\end{equation}
where the first order in $\chi$ cancels exactly \cite{supp}.
The maximal violation of Bell's inequalities is decreased by interaction, to the second order in $\chi$. 
It is possible to demonstrate  \cite{supp} that this result holds also for more than two wells. 
This suggests that interactions among particles should not be a key problem for the violation of Bell's inequalities in our system.

{\it Conclusions.}
Neutral atoms are currently playing a key role for the creation many-particles entanglement
\cite{EsteveNature08,RiedelNature10,GrossNature10,LuckeScience11,HaasSCIENCE2014}.
Yet, the investigation of nonlocality in these systems has been scarcely considered.   
In this manuscript we have proposed a scheme to create quantum nonlocality of an ultra cold gas trapped in an optical lattice, 
requiring only linear operations (balanced coupling between internal state of neighboring lattice sites), collective phase shifts,
and involving an arbitrarily large number of neutral atoms.
Nonlocality can be demonstrated via the violation of a novel set of Bell's inequalities recently 
proposed in \cite{TuraScience14} and involving only one- and two-body correlations.
For two wells, the protocol reduces to the one first proposed by Yurke and Stoler in \cite{YurkePRA92}.
For many wells, the relative violation goes as $1/N$ as a function of the number of atoms involved in the measurement. The violation of the Bell inequality
requires postselection, as typical \cite{YurkePRL92,MullinPRA08}.  
Our work paves the way toward the realization and demonstration of quantum nonlocality with ultracold atoms -- involving a large number of parties --
and their use for quantum information protocols.    

\begin{acknowledgements} 
{\it Acknowledgements.} We thank T. Wasak and S. Altenburg for useful discussions. 
This work has been partially supported by the Seventh Framework Programme for 
Research of the European Commission, under FET-Open grant QIBEC (No. 284584).
\end{acknowledgements}

\end{document}


\title{Supplementary Information: Nonlocality with ultracold atoms in a lattice}

\author{Sophie Pelisson}\affiliation{QSTAR, INO-CNR and LENS, Largo Enrico Fermi 2, I-50125 Firenze, Italy}
\author{Luca Pezz\'e}\affiliation{QSTAR, INO-CNR and LENS, Largo Enrico Fermi 2, I-50125 Firenze, Italy}
\author{Augusto Smerzi}\affiliation{QSTAR, INO-CNR and LENS, Largo Enrico Fermi 2, I-50125 Firenze, Italy}
\date{\today}

\maketitle

Here we report the calculation of the optimized Bell operator for the two- and three-wells case, including the effect of 
interactions between atoms during the balanced beam splitter used to couple neighboring wells in our system.
In particular, we show that the first order in the intensity of interaction exactly cancels. 
By symmetry, since our protocol involves coupling between neighboring wells, 
we expect that the result holds for any number of wells.

\section{two-wells case}
The interactions between atoms during the first beam splitter can be modelled by a non linear beam splitter as
\begin{equation}
\text{BS}_1=e^{-i\frac{\pi}{2}\left(J_y+\chi J_z^2\right)}
\label{NLBS}
\end{equation}
where $J_y=\frac{1}{2i}\left(c^\dag_{g,1}c_{e,2}-c^\dag_{e,2}c_{g,1}+c^\dag_{g,2}c_{e,1}-c^\dag_{e,1}c_{g,2}\right)$ and $J_z=\frac{1}{2}\left(c^\dag_{g,1}c_{g,1}-c^\dag_{e,1}c_{e,1}+c^\dag_{g,2}c_{g,2}-c^\dag_{e,2}c_{e,2}\right)$. \\
The coefficient $\chi$ represents the relative strength of the interaction with respect to the stimulated tunneling between wells. As 
\begin{equation}\begin{split}
e^{-i\hat{H}}\ket{\psi}&=\sum_ke^{-i\hat{H}}\ket{k}\bra{k}\psi\rangle\\
&=\sum_ke^{-i\lambda_k}\ket{k}\bra{k}\psi\rangle
\label{expop}\end{split}\end{equation}
with $\hat{H}\ket{k}=\lambda_k\ket{k}$, we then have to diagonalize the hamiltonian 
\begin{equation}
\hat{H}_i=\frac{\pi}{2}\left(J_y+\chi J_z^2\right)
\label{Hi}
\end{equation}
on the base of all permutations of 2 particles in two wells with 2 internal levels per well
\begin{equation}
\{\ket{2000},\ket{0200},\ket{0020},\ket{0002},\ket{1010},\ket{1001},\ket{0110},\ket{0101},\ket{1100},\ket{0011}\}.
\end{equation}

We can then write the state after the beam splitter as
\begin{equation}\begin{split}
e^{-i\hat{H}_i}\ket{\psi_0}&= \frac{e^{-i\frac{\chi\pi}{2}}}{2}\left(\ket{1010}+\ket{0101}\right)+2\left(\ket{0101}-\ket{1010}\right)\left(\frac{e^{-i\frac{\pi}{4}(\sqrt{4+\chi^2}-\chi)}}{4+\left(\chi+\sqrt{4+\chi^2}\right)^2}+\frac{e^{-i\frac{\pi}{4}(\sqrt{4+\chi^2}+\chi)}}{4+\left(\sqrt{4+\chi^2}-\chi\right)^2}\right)\\
&+i\left(e^{i\frac{\pi}{4}(\sqrt{4+\chi^2}-\chi)}\frac{\chi+\sqrt{4+\chi^2}}{4+\left(\chi+\sqrt{4+\chi^2}\right)^2}-e^{-i\frac{\pi}{4}(\sqrt{4+\chi^2}+\chi)}\frac{\sqrt{4+\chi^2}-\chi}{4+\left(\sqrt{4+\chi^2}-\chi\right)^2}\right)\left(\ket{1100}+\ket{0011}\right)
\end{split}\end{equation}
If we suppose that $\chi<<1$ then we can develop the former expression so that we obtain
\begin{equation}\begin{split}
e^{-i\hat{H}_i}\ket{\psi_0}&\simeq\frac{1}{2}e^{-i\frac{\pi\chi}{2}}\left(\ket{1010}+\ket{0101}\right)+\frac{i\chi}{4}e^{-i\frac{\pi\chi}{4}}\left(\ket{0101}-\ket{1010}\right)+\frac{1}{2}e^{-i\frac{\pi\chi}{4}}\left(\ket{1100}+\ket{0011}\right)\label{newini}
\end{split}\end{equation}
We can now come back to our sequence of local transformations and measurement and apply it to the new state \eqref{newini}. We will now describe this sequence involved in the calculation of $S_{00}$ as an example. First of all, we apply the phase shift on the ground atomic state:
\begin{equation}
\ket{\tilde{\psi}_{00}}=\frac{1}{2}e^{-i\pi\frac{\chi}{2}}\left(e^{i\left(\theta_1+\theta_2\right)}\ket{1010}+\ket{0101}\right)+i\frac{\chi}{4}e^{-i\pi\frac{\chi}{4}}\left(\ket{0101}-e^{i\left(\theta_1+\theta_2\right)}\ket{1010}\right)+\frac{1}{2}e^{-i\pi\frac{\chi}{4}}\left(e^{i\theta_1}\ket{1100}+e^{i\theta_2}\ket{0011}\right)
\end{equation}
Then, we apply the beam splitter in each well separately so that the state becomes
\begin{equation}\begin{split}
\ket{\psi_{00}}&=\left[\frac{1}{4}e^{-i\pi\frac{\chi}{2}}\left(e^{i\left(\theta_1+\theta_2\right)}+1\right)+i\frac{\chi}{8}e^{-i\pi\frac{\chi}{4}}\left(1-e^{i\left(\theta_1+\theta_2\right)}\right)\right]\left(\ket{1010}+\ket{0101}\right)\\
&+\left[\frac{1}{4}e^{-i\pi\frac{\chi}{2}}\left(e^{i\left(\theta_1+\theta_2\right)}-1\right)-i\frac{\chi}{8}e^{-i\pi\frac{\chi}{4}}\left(1+e^{i\left(\theta_1+\theta_2\right)}\right)\right]\left(\ket{1001}+\ket{0110}\right)\\
&+\frac{1}{2\sqrt{2}}e^{-i\pi\frac{\chi}{4}}\left[e^{i\theta_1}\left(\ket{0200}-\ket{2000}\right)+e^{i\theta_2}\left(\ket{0002}-\ket{0020}\right)\right]
\end{split}\end{equation}
As a consequence, the correlation $S_{00}$ is written as after selection of the useful part of the state
\begin{equation}\begin{split}
S_{00}&=2\bra{\psi_{00}}(-1)^{n_g^{(1)}+n_g^{(2)}}\ket{\psi_{00}}\\
&=2\left[\frac{e^{i\pi\frac{\chi}{2}}}{4}\left(e^{-i\left(\theta_1+\theta_2\right)}+1\right)-i\frac{\chi e^{i\pi\frac{\chi}{4}}}{8}\left(1-e^{-i\left(\theta_1+\theta_2\right)}\right)\right]\left[\frac{e^{-i\pi\frac{\chi}{2}}}{4}\left(e^{i\left(\theta_1+\theta_2\right)}+1\right)+i\frac{\chi e^{-i\pi\frac{\chi}{4}}}{8}\left(1-e^{i\left(\theta_1+\theta_2\right)}\right)\right]\\
&\qquad\qquad\left(\bra{1010}+\bra{0101}\right)\left(\ket{1010}+\ket{0101}\right)\\
&\quad+2\left[\frac{e^{i\pi\frac{\chi}{2}}}{4}\left(e^{-i\left(\theta_1+\theta_2\right)}-1\right)+i\frac{\chi e^{i\pi\frac{\chi}{4}}}{8}\left(1+e^{-i\left(\theta_1+\theta_2\right)}\right)\right]\left[\frac{e^{-i\pi\frac{\chi}{2}}}{4}\left(e^{i\left(\theta_1+\theta_2\right)}-1\right)-i\frac{\chi e^{-i\pi\frac{\chi}{4}}}{8}\left(1+e^{i\left(\theta_1+\theta_2\right)}\right)\right]\\
&\qquad\qquad\left(\bra{1001}+\bra{0110}\right)\left(-\ket{1001}-\ket{0110}\right)\\
&=\frac{1}{8}\left[4+\chi^2-\left(4-\chi^2\right)\cos\left(\theta_1+\theta_2\right)+4\chi\cos\frac{\pi\chi}{2}\sin\left(\theta_1+\theta_2\right)\right]\\
&\quad-\frac{1}{8}\left[4+\chi^2-\left(4-\chi^2\right)\cos\left(\theta_1+\theta_2\right)-4\chi\cos\frac{\pi\chi}{2}\sin\left(\theta_1+\theta_2\right)\right]\\
&=\frac{1}{4}\left[-2\left(4-\chi^2\right)\cos\left(\theta_1+\theta_2\right)+8\chi\cos\frac{\pi\chi}{2}\sin\left(\theta_1+\theta_2\right)\right]
\end{split}\end{equation}
Proceeding the same for the terms $S_{11}$ and $S_{01}$, we finally obtain the new expression for the Bell's operator 
\begin{equation}\begin{split}
B(\theta_1,\theta_2,\theta'_1,\theta'_2)&=\frac{-4+\chi^2}{4}\left[\cos\left(\theta_1+\theta_2\right)-\cos\left(\theta_1'+\theta_2'\right)+\cos\left(\theta_1'+\theta_2\right)+\cos\left(\theta_1+\theta_2'\right)\right]\\
&+\chi\cos\frac{\pi\chi}{2}\left[\sin\left(\theta_1+\theta_2\right)-\sin\left(\theta_1'+\theta_2'\right)+\sin\left(\theta_1'+\theta_2\right)+\sin\left(\theta_1+\theta_2'\right)\right]
\label{Bop}
\end{split}\end{equation}
To better understand the role of the inter-atomic interactions on the violations of Bell's inequalities, we look at what happens to the expression \eqref{Bop} when we set the optimal $\theta_i,\theta_i'$ reported in the main text. With these values, we then obtain
\begin{equation}
|B_\text{opt}|=\frac{4-\chi^2}{\sqrt{2}}=2\sqrt{2}-\frac{\chi^2}{\sqrt{2}}
\end{equation}
For $\chi<<1$ this expression is always greater than 2 and violates the Bell's inequalities. 

\section{three-wells case}
\subsection{Non-linear beam splitter and reduction of the base}

We proceed the same way for the 3 well case reusing the expression \eqref{NLBS} with
\begin{equation}\left\{\begin{array}{ll}
J_y=\frac{1}{2i}\left(a^\dag_{g,1}a_{e,3}-a^\dag_{e,3}a_{g,1}+a^\dag_{g,2}a_{e,1}-a^\dag_{e,1}a_{g,2}+a^\dag_{g,3}a_{e,2}-a^\dag_{e,2}a_{g,3}\right)&\\
J_z=\frac{1}{2}\left(a^\dag_{g,1}a_{g,1}-a^\dag_{e,1}a_{e,1}+a^\dag_{g,2}a_{g,2}-a^\dag_{e,2}a_{e,2}+a^\dag_{g,3}a_{g,3}-a^\dag_{e,3}a_{e,3}\right)&
\end{array}\right.\end{equation}
Similarly to the 2 wells case, we have to diagonalize the hamiltonian 
\begin{equation}
\hat{H}_i=\frac{\pi}{2}\left(J_y+\chi J_z^2\right)
\label{Hi}
\end{equation}
on the base of all permutations of 3 particles in 3 wells with 2 internal levels per well. However, this base is huge and difficult to deal with analytically. In order to reduce the size of this basis, we notice that our initial state is very simple as it presents only one atom per well in its ground state, $\ket{\psi_0}=\ket{101010}$, and that only the $J_y$ part of the beam splitter is mixing the states, we can then reduce the base to the states arising from the linear beam splitter when applied to our ground states so that to the eight states below
\begin{equation}
\{\ket{101010},\ket{001011},\ket{110010},\ket{010011},\ket{101100},\ket{010101},\ket{001101},\ket{110100}\}.
\end{equation}
Writing the Hamiltonian \eqref{Hi} in this reduced base and diagonalizing it allows to find the effect of our non linear beam splitter on the initial state.
\begin{equation}\begin{split}
\ket{\psi}=&e^{-i\hat{H}_i}\ket{\psi_0}\\
=&\frac{(1+i)e^{-i\pi\chi/4}}{2\sqrt{2(16+\chi(\chi-4))}}\left\{\right.\\
&\left[e^{i\frac{\pi}{8}\sqrt{16+\chi(\chi-4)}}(2-\chi+\sqrt{16+\chi(\chi-4)})+e^{-i\frac{\pi}{8}\sqrt{16+\chi(\chi-4)}}(2-\chi-\sqrt{16+\chi(\chi-4)})\right]\left(\ket{101010}+i\ket{010101}\right)\\
&\left.+2\cos\left(\frac{\pi}{8}\sqrt{16+\chi(\chi-4)}\right)\left[\ket{010011}+\ket{001101}+\ket{110100}+i(\ket{001011}+\ket{110010}+\ket{101100})\right]\right\}\\
&\frac{(1-i)e^{-i\pi\chi/4}}{2\sqrt{2(16+\chi(\chi+4))}}\left\{\right.\\
&\left[e^{i\frac{\pi}{8}\sqrt{16+\chi(\chi+4)}}(2+\chi-\sqrt{16+\chi(\chi+4)})+e^{-i\frac{\pi}{8}\sqrt{16+\chi(\chi+4)}}(2+\chi+\sqrt{16+\chi(\chi+4)})\right]\left(\ket{101010}-i\ket{010101}\right)\\
&\left.+2\cos\left(\frac{\pi}{8}\sqrt{16+\chi(\chi+4)}\right)\left[\ket{010011}+\ket{001101}+\ket{110100}-i(\ket{001011}+\ket{110010}+\ket{101100})\right]\right\}
\label{BS1}\end{split}\end{equation}

\subsection{Measurement sequence}

Once we have the exit state from the first beam splitter, we can apply the set of local transformation and measurement, i.e. the phase shift and the second beam splitter in each well separately before the measurement of the parity of the number of particles in ground state. We have here to recall that in order to measure a violation of the Bell's inequalities, we have to post select the states where all the wells are populated. As the second beam splitter is applied in each well separately, we can select the state directly from \eqref{BS1}. As a consequence, the state $\ket{\psi}$ can be rewritten as 
\begin{equation}
\ket{\psi}=\frac{1}{\sqrt{\mathcal{N}(\chi)}}\left[\left(F1(\chi)+F2(\chi)\right)\ket{101010}+i\left(F1(\chi)-F2(\chi)\right)\ket{010101}\right]
\label{psi}\end{equation}
Where $\mathcal{N}(\chi)$ represents the normalization function of the selected state and
\begin{equation}\left\{\begin{array}{ll}
F1(\chi)=\frac{(1+i)e^{-i\pi\chi/4}}{2\sqrt{2(16+\chi(\chi-4))}}\left[e^{i\frac{\pi}{8}\sqrt{16+\chi(\chi-4)}}(2-\chi+\sqrt{16+\chi(\chi-4)})+e^{-i\frac{\pi}{8}\sqrt{16+\chi(\chi-4)}}(2-\chi-\sqrt{16+\chi(\chi-4)})\right]&\\
F2(\chi)=\frac{(1-i)e^{-i\pi\chi/4}}{2\sqrt{2(16+\chi(\chi+4))}}\left[e^{i\frac{\pi}{8}\sqrt{16+\chi(\chi+4)}}(2+\chi-\sqrt{16+\chi(\chi+4)})+e^{-i\frac{\pi}{8}\sqrt{16+\chi(\chi+4)}}(2+\chi+\sqrt{16+\chi(\chi+4)})\right]&
\end{array}\right.\end{equation}

We want to investigate the behavior of the Bell operator with the first order of $\chi$. So we rewrite the coefficient in \eqref{psi} in series of $\chi$ as
\begin{equation}\left\{\begin{array}{ll}
F1(\chi)-F2(\chi)\simeq-i\sqrt{2}+\frac{9\pi\chi}{16\sqrt{2}}+\mathcal{O}(\chi^2)&\\
F1(\chi)+F2(\chi)\simeq\sqrt{2}+\frac{9i\pi\chi}{16\sqrt{2}}+\mathcal{O}(\chi^2)&
\end{array}\right.\end{equation}
and $\mathcal{N}(\chi)\simeq4+\frac{9\pi^2\chi^2}{128}$. Using these expressions to rewrite \eqref{psi} and applying to this state the sequence of measurement in order to observe Bell's inequalities we can deduce the expression of the Bell's operator modified by interactions in first order of $\chi$ as
\begin{equation}
\tilde{B}_3(\theta,\varphi)=\frac{2}{\mathcal{N}(\chi)}\left(3+\frac{81\pi^2\chi^2}{512}\right)B_3(\theta,\varphi)
\label{Bmod}\end{equation}
And the coefficient in \eqref{Bmod} is equal to 1 showing that the first order in $\chi$ is cancelled for 3 wells. As a consequence, symmetry of the system allows us to conclude that for any number of wells, the first order of interaction is always cancelled.